\documentclass[a4paper,12pt]{article}
\pdfoutput=1 

\usepackage{jheppub} 
\usepackage{hyperref}
\hypersetup{
    bookmarks=true,         
    unicode=false,          
    pdftoolbar=true,        
    pdfmenubar=true,        
    pdffitwindow=false,     
    pdfstartview={FitH},    
    pdftitle={My title},    
    pdfauthor={Author},     
    pdfsubject={Subject},   
    pdfcreator={Creator},   
    pdfproducer={Producer}, 
    pdfkeywords={keyword1} {key2} {key3}, 
    pdfnewwindow=true,      
    colorlinks=true,       
    linkcolor=red,          
    citecolor=purple,        
    filecolor=magenta,      
    urlcolor=blue,           
    linktocpage=true
}
\usepackage{graphics}
\usepackage{rotating}
\usepackage{subfigure}
\usepackage{pstricks}
\usepackage{color}
\usepackage{amsfonts}
\usepackage{mathrsfs,amsmath}
\usepackage{epsfig}
\usepackage{amsmath,amssymb,amsthm,graphicx,latexsym}
\usepackage[toc,page]{appendix}

\usepackage{ulem}

\newcommand{\be}{\begin{equation}}
\newcommand{\ee}{\end{equation}}
\newcommand{\bea}{\begin{eqnarray}}
\newcommand{\eea}{\end{eqnarray}}
\newcommand{\bml}{\begin{subequations}}
\newcommand{\eml}{\end{subequations}}
\newcommand{\bfig}{\begin{figure}}
\newcommand{\efig}{\end{figure}}

\newcommand{\bmat}{\begin{pmatrix}}
\newcommand{\emat}{\end{pmatrix}}
\begin{document}
	$~~~~~~~~~~~~~~~~~~~~~~~~~~~~~~~~~~~~~~~~~~~~~~~~~~~~~~~~~~~~~~~~~~~~~~~~~~~~~~~~~~~~$
	\title{\textsc{\fontsize{25}{17}\selectfont \sffamily \bfseries {Generic 3-point Statistics with Tensor Modes in Light of Swampland and TCC}}}
	
	\author{Abhishek Naskar,}			
	\author{Supratik Pal}
	\affiliation{Physics and Applied Mathematics Unit, Indian Statistical Institute, 203 B.T. Road,
	Kolkata  700108, India}
	\emailAdd{abhiatrkmrc@gmail.com, supratik@isical.ac.in}

	\abstract{Recently proposed Swampland Criteria (SC) and Trans-Planckian Censorship Conjecture (TCC) together put stringent
	theoretical constraints on slow roll inflation, raising a  question
	on future prospects of detection of Primordial Gravitational Waves (PGW). As it appears, the only option to relax
	the constraints  is by considering Non Bunch Davies (NBD) initial states, that in turn brings back the observational relevance of PGW
	via its 2-point function. In this article we develop consistent 3-point statistics with tensor modes  for all possible correlators (auto and mixed)
	for NBD initial states in the
	light of SC and TCC in a generic, model independent 
	framework of Effective Field Theory of inflation. 
	We also construct the templates of the corresponding nonlinearity parameters $f_{NL}$ for  different shapes of relevance
	and investigate if any of the 3-point correlators could be  of interest for future CMB missions. Our analysis reveals
	that the prospects of detecting the tensor auto correlator are almost nil whereas the mixed correlators might be relevant for future CMB missions.
	}


	\maketitle
	\flushbottom
	\section{Introduction}

Nowadays any viable model of inflation has to pass through two crucial hurdles: one, the Swampland Criteria (SC)  \cite{Agrawal:2018own}, 
and the other, the Trans-Planckian Censorship Conjecture (TCC) \cite{Bedroya:2019tba}. 
 Even though 	
	the success of Inflationary Cosmology lies in its indigenous  ability to solve the puzzles with 
 Standard Big Bang Cosmology and to provide the seeds for Large Scale Structures (LSS) we see today,
 as well as in its profound consistency with the highly precise Cosmic Microwave Background (CMB) data, the latest being the Planck 2018 data  \cite{Akrami:2018odb}, recently it was revealed that  the single field, slow roll inflationary paradigm  itself is plagued with serious theoretical inconsistency 
  \cite{Agrawal:2018own}.  In brief, in order to 
  embed any model of inflation consistently in a Quantum Gravity theory, it   has to satisfy
a set of criteria, collectively called  the Swampland Criteria, that states
  that  
 $(i)$ the field excursion by scalar fields in field space is bounded by $\Delta \phi < \mathcal{O}(1)M_{pl}$;  and $(ii)$ the potential of a scalar field which rolls and dominates the energy density of the universe
 has to satisfy the condition $\frac{V'(\phi)}{V(\phi)}\sim\sqrt{2\epsilon}>c M_{pl}^{-1}$ 
 with $c\sim\mathcal{O}(1)$ and
 $\epsilon$ being the first slow roll parameter and $M_{pl}$ the Planck mass. However, as is well-known, for slow roll inflation, based on which many inflationary models have been proposed,
 the potential of inflaton field has to be sufficiently flat with the current observational bound for $\epsilon < 0.0063$ \cite{Akrami:2018odb}, which is nowhere close to $\mathcal{O}(1)$. This is in direct contradiction with the second
SC, thereby putting   most of the existing inflationary models in tension.
 
The second constraint on the theory of inflation comes from recently conjectured Trans-Planckian Censorship Conjecture \cite{Bedroya:2019tba}. In order to produce
 the seeds of perturbations in LSS, the quantum vacuum fluctuations
  exit the horizon and get frozen during inflation
only to re-enter the horizon at a later stage of cosmological evolution. However, one can see that if inflation lasts more than the minimum period required, all the observable modes has to have a length scale smaller than Planck scale, which once again leads to
 inconsistent theory of Quantum Gravity.  This problem is known as Trans-Planckian Problem 
 \cite{Martin:2000xs}. The only wayout seems to be coming through a conjecture, called the Trans-Planckian Censorship Conjecture \cite{Bedroya:2019tba}, 
 that states that the length scales
 smaller than Planck scale can never exit the Hubble Horizon. As a consequence this conjecture
 places a tight theoretical constraint on the first slow roll parameter as $\epsilon<10^{-31}$. 
 
 Looking at the constraint on $\epsilon$ put by the SC and TCC, it may appear  that they are in tension with each other, and together they
  may serve as a no go theorem for
 a consistent
 slow roll model of inflation. 
 However,  it was revealed of late that TCC can be derived from the first SC \cite{Brahma:2019vpl}. Also the second
 SC can be refined using the first SC and Bousso's entropy bound \cite{Bousso:1999xy} and 
  can be restated as,
  a field theory will not be in Swampland if either $\frac{V'(\phi)}{V(\phi)}\sim\sqrt{2\epsilon}>c M_{pl}^{-1}$ or $\frac{V''(\phi)}{V(\phi)}\leq c' M_{pl}^{-2}$ with
  $c,c'\sim \mathcal{O}(1)$  \cite{Ooguri:2018wrx}. 
  Consequently, one can have a valid inflationary scenario consistent with the refined SC if one satisfies the following condition:
 the second slow roll parameter 
  $\eta\sim \mathcal{O}(1)$ with the first slow roll parameter still satisfying the observational bound $\epsilon<<1$. The constraint on $\eta$ from Planck observation 
  \cite{Akrami:2018odb}  is given by $\eta\sim -0.02$ 
which can be considered as close to $\mathcal{O}(1)$ as argued in \cite{Brahma:2020cpy} . As we have already stated, the only requirement from TCC is $\epsilon \ll 1$.
Hence, to summaries the theoretical constraints, 
any slow roll inflationary theory would be consistent  with both the  refined SC
and TCC if one satisfies  $\epsilon<<1$ and $\eta\sim \mathcal{O}(1)$ simultaneously.
 
 Even though one takes care of the   theoretical perspective and the possible wayout as discussed above while inflationary 
 model building, on the backdrop
this leads to an altogether novel  problem from observational point of view, namely, the future prospects of detection of primordial gravitational waves (PGW).
As is well-known, any constraint on the first slow roll parameter $\epsilon$ transmits to a corresponding constrain on the PGW amplitude, 
namely, the  tensor-to-scalar ratio $r$  via
 the single field consistency relation $r=16 \epsilon$ 
 \cite{Maldacena:2002vr}.
The old SC requires that $r>1$ which is ruled out by observation with the current upper bound 
from Planck 2018 being $r<0.064$ \cite{Akrami:2018odb}. Though the 
 refined SC as stated above is consistent with the current bound,
TCC requires a very tiny value for  $r<10^{-30}$. This tiny value is beyond the scope of any future observation, let alone the next generation CMB missions like CMB-S4 
\cite{Abazajian:2016yjj},  LiteBIRD \cite{Matsumura:2013aja,Suzuki:2018cuy}, 
COrE \cite{Delabrouille:2017rct} etc.
This bound is further tightened  to $r<10^{-47}$ if one
  assumes that the pre-inflationary era was
 radiation dominated \cite{Brandenberger:2019eni}. Although the bound  can slightly be relaxed to $r<10^{-10}$
 if one assumes that the equation of state of inflation changes from $-1$ to $-\frac{1}{3}$ after a 
 couple of e-folding \cite{Kamali:2019gzr}, this is still much
below the scope of future CMB missions.
 Thus,  those theoretical constraints put  a serious question on the prospects of PGW to act as 
 the smoking gun of inflationary paradigm 
 \cite{Shiraishi:2019yux}, namely, whether or not  it can act  as the  'Holy Grail' of inflationary cosmology in the 
 light of SC and TCC.

 
However, with further progress of 
 theoretical analysis, it was 
 recently found that one can bypass the SC   within 
 the standard  single field 
 inflationary framework if one considers the initial states as follows:  the tensor modes in 
 Non Bunch Davies state (NBD) and scalar modes in  Bunch Davies (BD) state  \cite{Brahma:2018hrd,Ashoorioon:2018sqb}. Subsequently, it
  was also revealed that  
 if the initial states of both the scalar and tensor modes are NBD, one can relax the TCC
 bound on $r$ as well, thereby revitalising the possibility of producing a detectable PGW by vacuum fluctuation of inflaton  \cite{Brahma:2019unn}
 orders close to the range of future CMB missions. Thus, the above prescription 
 of considering NBD states for both scalar and tensor modes to obtain a detectable PGW signal from a consistent theoretical 
 framework brings back the relevance of  PGW as a probe of inflationary
cosmology. 
The effect of NBD on the scalar perturbations has been studied to a considerable extent
 \cite{Holman:2007na,Agullo:2010ws,Kundu:2011sg,Ganc:2011dy,Flauger:2013hra,Chandra:2016jll}, mostly with theoretical motivation.
 However, in the light of PGW,  the motivation to explore NBD states gets an observational relevance as well.

 In this article our primary intention is to develop   consistent 3-point statistics with tensor modes with NBD initial states  keeping
 in mind the above discussion on its relevance in the light of  SC and TCC.
 To this end, we would bring forth different correlators relevant for the non-Gaussian behaviour of PGW followed by the templates for the corresponding nonlinearity parameters $f_{NL}$ for different relevant shapes 
 considering NBD initial states for both scalar and tensor modes; and study their prospects of detection
 in future CMB missions.
We believe the present analysis  is important for couple of reasons. First, the higher order correlation functions can be very sensitive to the choice of 
 initial state (BD/NBD) and hence can be an important probe for the initial vacuum as well. 
 Secondly,   the shape of the auto correlation of tensor modes can shed some light on the source of 
 PGW. Thirdly, the mixed correlator of scalar and tensor modes can probe additional properties of PGW
 which will be discussed in the next sections. To name one, correlators
 of tensor-scalar-scalar type  can serve as a probe of spatial diffeomorphism breaking during inflation 
 \cite{Bartolo:2015qvr}. From observational point of view, this mixed correlator also reflects on the quadrupole moment and 
might be of interest for future CMB missions. 
Last but not the least, these non-Gaussian properties serve as
 additional probe for PGW along with the 2-point function $r$. 
 Thus, detection of any of the correlators would give us a hint on PGW and initial state.
 Some preliminary studies on the prospects
  of tensor non-Gaussianities have been reported  by the present authors in
   \cite{Naskar:2018rmu,Naskar:2019shl}
  and also by some other authors
  \cite{Shiraishi:2019yux,Namba:2015gja,aniket,Dimastrogiovanni:2018gkl}.
  However, almost all of them consider BD vacuum. Here we would like to extend the analysis for NBD states and explore the prospects
  of both auto and mixed correlators.
  
 As of now,  the
  observational bounds on tensor  non-Gaussianities 
are  not too tight.  The current constraints on the amplitude of tensor bispectrum in equilateral limit is given by Planck 2018 as:
 $f_{NL}^{T} = 800 \pm 1100$ \cite{Akrami:2019izv}. 
However, for the reasons mentioned above, the hunt for any  possible non-Gaussian behaviour of PGW in upcoming CMB missions will be more relevant than ever.
These constraints will be further improved in next generation
 CMB missions.  For example, LiteBIRD \cite{Matsumura:2013aja,Suzuki:2018cuy} 
 targets to improve the constraint of the amplitude of tensor bispectra by three orders of magnitude 
 \cite{Matsumura:2013aja, Suzuki:2018cuy}. The constraint on tensor-tensor-scalar  mixed correlator will be improved by CMB-S4 
  \cite{Abazajian:2016yjj}. Others have their own specific target.
Keeping this in mind,  in this article, we would like to explore all possible correlators, namely, the  tensor-tensor-tensor (auto) correlator and
 tensor-tensor-scalar and  tensor-scalar-scalar (both mixed) in the light of SC and TCC. 
  The theoretical analysis is done using the model independent framework of 
  Effective Field Theory (EFT) of inflation \cite{Cheung:2007st} to calculate
  the tensor-tensor-tensor and scalar-tensor-tensor correlators. To calculate the
  tensor-scalar-scalar correlator  we have used the
  EFT of inflation with broken space-time diffeomorphism \cite{Bartolo:2015qvr} as this type of correlator can
  reflect a unique feature of broken spatial diffeomorphism. Once we have the correlators in our hand, we would move on to proposing templates for the corresponding nonlinearity parameters $f_{NL}$ for different shapes of interest
and would also investigate if any of them could be the point of interest for future CMB missions by finding out the possible upper limits for each one allowed by the parameters of the theory under consideration. We would like to reiterate that because of the background EFT of inflation, our calculations and results are more or less generic and model-independent. 
 

 
\section{Swampland, TCC and Non Bunch Davies states}\label{sec2}

As already mentioned, it is the non Bunch Davies states that can give rise to a viable inflationary scenario respecting the SC and TCC.
Let us begin our discussion with 
a brief review  on the role of NBD states and how that can help  bypass the SC and  relax the bound
on $r$ coming from TCC in the same vein  of \cite{Brahma:2018hrd,Ashoorioon:2018sqb,Brahma:2019unn}. This will also help
us develop the rest of the article from a consistent theoretical setup. To this end, we will first discuss the scenario with the minimal model,
i.e., single field inflation with canonical kinetic term, and will subsequently move on to discussing the non-canonical inflationary models.

\subsection{Canonical models of inflation}
For a canonical inflation scenario the power spectra of scalar and tensor modes for NBD
states are given by
\begin{equation}\label{eq2.1}
P_{\zeta}(k)= \frac{H^2}{8 \epsilon M_{pl}^2}|\alpha^{s}_k+\beta^{s}_k|^2
\end{equation}
and,
\begin{equation}\label{eq2.2}
P_{\gamma}(k)= \frac{2 H^2}{M_{pl}^2}|\alpha^{t}_k+\beta^{t}_k|^2
\end{equation}
where $\zeta$ and $\gamma$ represent scalar and tensor modes respectively; $\alpha$ and $\beta$'s are Bogolyubov coefficients and the indices 
$s$ and $t$ signify
scalar and tensor respectively as we have considered different vacua for scalar and tensor modes. 
Setting $\alpha^{s/t}_k=1$ and $\beta^{s/t}_k=0$ one readily gets
back the BD states. 

As a result the tensor-to-scalar ratio can be written as,
\begin{equation}\label{eq2.3}
r=\frac{P_{\gamma}(k)}{P_{\zeta}(k)}=16 \epsilon \Gamma
\end{equation}
where $\epsilon$ is the first slow roll parameter and
\begin{equation}\label{eq2.4}
\Gamma=\frac{|\alpha^{t}_k+\beta^{t}_k|^2}{|\alpha^{s}_k+\beta^{s}_k|^2}
\end{equation}

The first theoretical constraint on the Bogolyubov coefficients come via the Wronskian condition
\begin{equation}
|\alpha^{s/t}_k|^2-|\beta^{s/t}_k|^2=1
\end{equation} 
using which the Bogolyubov coefficients can  further be parametrized as
\begin{eqnarray}\label{2.6}
\alpha^{s/t}_k&=&\sqrt{1+N_k^{(s/t)}} e^{i \theta_{\alpha}^{(s/t)}(k)}\\ \label{2.7}
\beta^{s/t}_k&=&\sqrt{N_k^{(s/t)}}e^{i\theta_{\beta}^{(s/t)}(k)}
\end{eqnarray}

Here, $N_k^{(s/t)}$ represents the number of NBD particles in BD state and $\theta_{\alpha}^{(s/t)}(k)$ and 
$\theta_{\beta}^{(s/t)}(k)$ are the phase factors. Consequently, one can define the relative phase as
\begin{equation}
\theta^{(s/t)}(k)=\theta_{\alpha}^{(s/t)}(k)-\theta_{\beta}^{(s/t)}(k)
\end{equation}
This parametrisation would help us to reduce the number of parameters since the individual phase factors are degenerate.

Further, there is a theoretical constraint on  the parameters $\beta^{(s/t)}_k$  coming from backreaction condition 
\cite{Holman:2007na,Greene:2004np}. For this one needs to model them as 
 \begin{equation}
\beta^{(s/t)}_k \sim \beta^{(s/t)}_0 e^{-\frac{k^2}{(M_{(s/t)}a(\eta_0))^2}}
\end{equation}
where $\eta_0$ describes the time when modes are below cut-off scale $M_{(s/t)}$. One can readily check  that 
for $k > M_{(s/t)}$, $\beta^{(s/t)}\rightarrow 0$. Now, in order a theory of inflation to be valid within the 
regime of EFT, $ M_{(s/t)}>H$. For slow roll inflation this condition translates into the following:
\begin{equation}\label{eq2.10}
\beta_0^{(s/t)}\leq \sqrt{\epsilon \eta'} \frac{H M_{pl}}{M_{(s/t)}^2}
\end{equation}
where $\eta'$ represents the second slow roll parameter.

As mentioned earlier, the  SC  forces the first slow roll parameter
$\epsilon\sim \frac{1}{2}\left(\frac{V'(\phi)}{V(\phi)}\right)^2\sim \frac{c^2}{2}$ to be $\mathcal{O}(1)$ 
which directly contradicts the slow roll condition of inflation and as a result the single 
field consistency relation now violates the observational bound on $r$. From \eqref{eq2.3} and 
\eqref{eq2.4} one
can see that for $\Gamma<1$ the observational constraint on $r$ can still be satisfied even with
a large $\epsilon$ coming from SC. Now there are two ways to make $\Gamma<1$.
The first one is to have NBD states for scalar fluctuations and BD state for tensor fluctuation.
One can choose a model where $\Gamma_s\left(=|\alpha^{s}_k+\beta^{s}_k|^2\right)>>1$ and suppress the 
value of $r$, without breaking the backreaction condition \eqref{eq2.10} as shown in \cite{Holman:2007na}.
But as explained in \cite{Ashoorioon:2018sqb} this choice of $\Gamma_s$ can ruin the constraint on 
scalar three point function. On the other hand, if one starts with tensor modes in NBD state and scalar modes in BD state
one can find a region in the parameter space that satisfy the PGW bispectrum constraints and allow
$c\sim 0.8$; larger value of $c\sim 0.9$ can also be achievable in this region of parameter space 
\cite{Brahma:2018hrd} thus satisfying the SC.

Secondly,  TCC puts  a very tight constraint on the first slow roll parameter 
$\epsilon<10^{-31}$ and hence from single field consistency relation for BD vacuum, $r<10^{-30}$. 
This bound on the tensor to scalar ratio basically states that in 
future if any PGW gets detected that can not be due to inflation but can be produced from other 
sources during inflation \cite{Namba:2015gja,aniket,Dimastrogiovanni:2018gkl,Naskar:2019shl}. 
So, apparently,  PGW loses its 'Holy Grail' status  in the light of TCC \cite{Brandenberger:2011eq}.
However,  this theoretical bound on $r$ placed by TCC can 
be relaxed significantly if one uses NBD states for both tensor and scalar modes \cite{Brahma:2019unn}. 
Consequently, the consistency relation now modifies to \eqref{eq2.3},
\begin{equation}
r=16 \epsilon \frac{\Gamma_t}{\Gamma_s}
\end{equation}
where the $\Gamma$ in \eqref{eq2.3} factor is re-written in terms of $\Gamma_{s/t}$ defined earlier in
this section. In this scenario TCC places the bound not on $r$ alone but on the combined parameter
 $\frac{\epsilon}{\Gamma_s}<10^{-31}$. So if
one can enhance the $\Gamma_t$ then the bound on $r$ can be relaxed significantly. To do so we take a look at the 
backreaction condition \eqref{eq2.10}. If one uses \eqref{eq2.1} to properly replace the
 $M_{pl}$ factor
of \eqref{eq2.10} one arrives at
\begin{equation}
\beta_0^{(t)}\leq \frac{\sqrt{\eta'}}{8 \pi^2 \mathcal{P_{\zeta}}} \left(\frac{H}{M_{(t)}}\right)^2 \sqrt{\Gamma_s}
\end{equation}

So with $\beta_0^{(s)}>>1$ we can have a large $\beta_0^{(t)}$. In this limit of $\beta_0^{(s)}$, 
$\sqrt{\Gamma_s}\sim \beta_0^{(s)}$. With $\mathcal{P}_{\zeta}\sim 10^{-9}$, $\eta'\sim -0.01755$ \cite{Akrami:2018odb}
if one considers, $\Gamma_s\sim 10^{23}$ with $M_{(t)}=10 H$ we can have $r \leq 0.001$ which is still within the detectable
range of next generation CMB missions. Here one important thing to note is that $\beta_0^{(s)}>>1$ does not spoil the
scalar non-Gaussianity bound as TCC requires a very small value of $\epsilon$ that guarantees that the 
bispectrum amplitude remains within the observational bound as it is evident from \cite{Ashoorioon:2013eia,Ashoorioon:2015pia}.
\begin{equation}
f_{NL}^{loc}\sim \frac{k_s}{k_l} \epsilon
\end{equation}
were $f_{NL}^{loc}$ is the amplitude of local configuration of scalar bispectrum and $k_s$ and $k_l$ are respectively the shortest and longest 
wavelength probed by any CMB mission under consideration (e.g., Planck mission as of today).


\subsection{Non-canonical models of inflation}
Until now we have only discussed the scenario with canonical kinetic term. However, EFT of inflation can take into account inflationary models with
non-canonical terms as well. 
 This will modify the consistency relation 
\cite{Garriga:1999vw} within the single field framework as well as  the backreaction condition \cite{Flauger:2013hra}. Effectively, the modifications come via 
the scalar and tensor  sound speeds $c_{s/t}$, including which \eqref{eq2.1} and \eqref{eq2.2} now look
\begin{eqnarray}
P_{\zeta}(k)&=& \frac{H^2}{8 \epsilon M_{pl}^2 c_s}\Gamma_s\label{eq2.14}\\
P_{\gamma}(k)&=& \frac{2 H^2}{M_{pl}^2 c_t}\Gamma_t \label{eq2.15}
\end{eqnarray}
Consequently,, the consistency relation now takes the form
\begin{equation}\label{eq2.16}
r=16 \epsilon \frac{\Gamma_t}{\Gamma_s}\frac{c_s}{c_t}
\end{equation}
and the backreaction condition modifies to
\begin{equation}\label{eq2.17}
\beta_0^{(s/t)}\leq \sqrt{\epsilon \eta'} \frac{H M_{pl}}{M_{(s/t)}^2}\frac{1}{\sqrt{c_{s/t}}}
\end{equation}

It is straightforward to check that for $c_{s/t} =1$ one gets back the canonical scenario.
From \eqref{eq2.16} we can see in the context of SC, apart from the $\frac{\Gamma_t}{\Gamma_s}$ term,
there is another factor $\frac{c_s}{c_t}$ which can also act as a suppression factor for a small  value for $c_s$.
However,  $c_s$ can not be arbitrarily small  as there is another constraint on it coming from  the bounds on scalar non-Gaussianities.
In the equilateral limit, the amplitude of scalar bispectrum  reads \cite{Kinney:2018nny, Chen:2006nt},
\begin{equation}
f_{NL}^{eq}=\frac{35}{108}\left(-1+\frac{1}{c_s^2}\right)
\end{equation}
Hence, together with the constraint from scalar non-Gaussianity  $f_{NL}^{eq}=-4\pm40$ \cite{Ade:2015ava}, the value of factor
$c$ coming from SC becomes $c\sim 0.37$ with $c_s>0.067$. 

Clearly, the above  value of $c$ is not 
compatible with SC, as discussed earlier. So,  in order to bypass the SC,
 one need take shelter of the NBD factor $\frac{\Gamma_t}{\Gamma_s}$.
However, for TCC  and refined SC 
this non-trivial sound speed can help in relaxing the bound on $r$. From \eqref{eq2.17} one can write
\begin{equation}
\beta_0^{(t)}\leq \frac{\sqrt{\eta'}}{8 \pi^2 \mathcal{P_{\zeta}}} \left(\frac{H}{M_{(t)}}\right)^2 \sqrt{\frac{\Gamma_s}{c_s c_t}}
\end{equation}

Together with \eqref{eq2.16} this equation tells us that in order to relax the bound on $r$, one needs to have a small value for the sound speed of
tensor fluctuation such that the factor $\frac{c_s}{c_t}$ acts as an enhancing factor.
\begin{figure}[htp] \label{fig1} 
\centering
  \includegraphics[width=.6\linewidth]{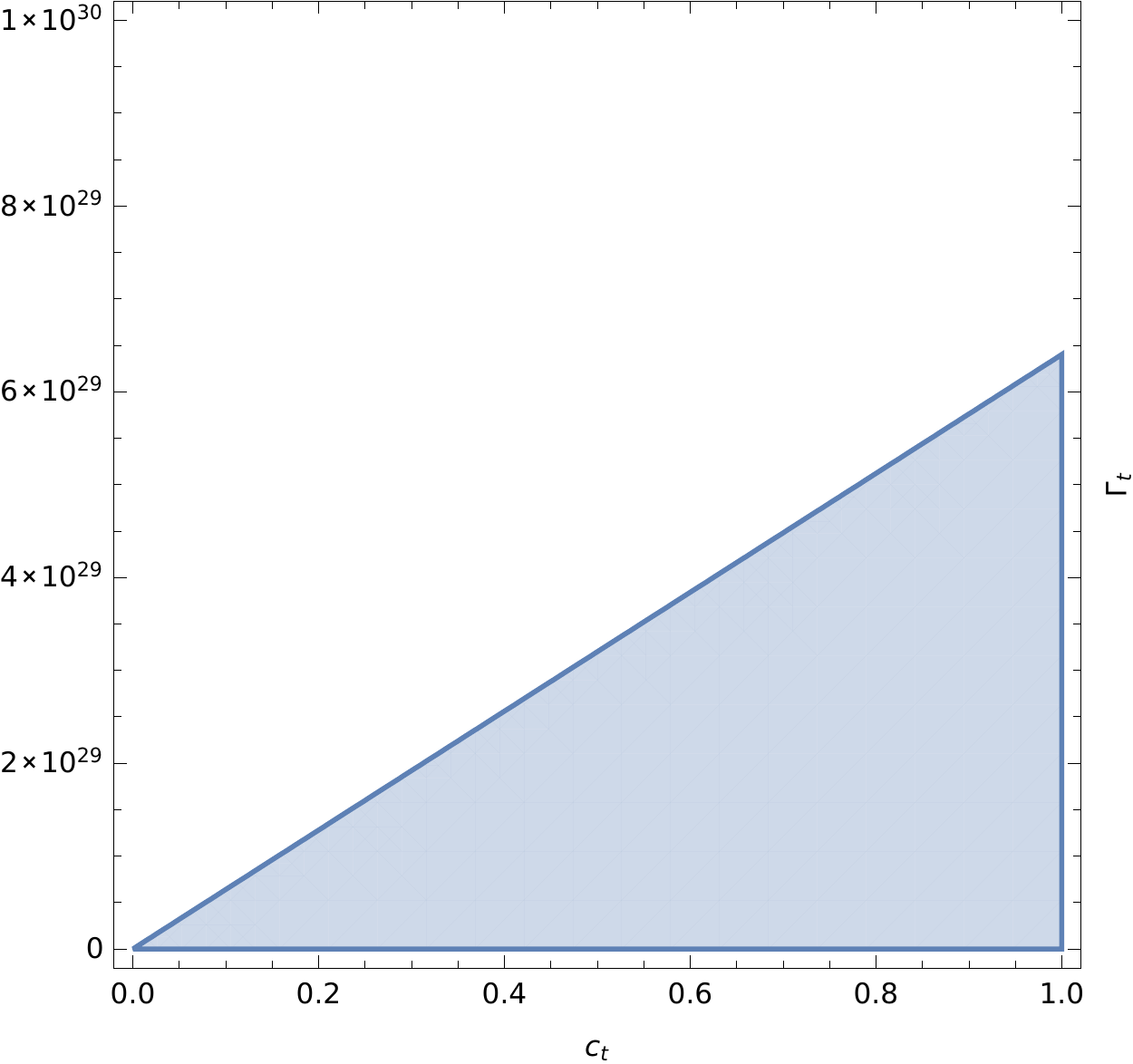}
\caption{Allowed region of $N_0^{(t)}$ and $c_t$ using Planck constraint on the amplitude of two point function.}
\end{figure}%

Combining all the above constraints on the parameters, we have $\frac{\epsilon c_s}{\Gamma_{(s)}}<10^{-31}$ and $r<0.064$. 
Fig \ref{fig1} show the 
region between $c_t$ and $N_0^{(t)}$ that is allowed considering both the theoretical (SC and TCC) and latest observational constraints
(Planck 2018). The figure show that for particular combinations of $c_t$ and 
$N_0^{(t)}$ one can have a bound on $r$ which can match the upper bound set by Planck 2018, thereby bringing back its relevance
as the smoking gun for Inflationary Cosmology.


 \section{The second order action from EFT}\label{sec3}
 
 In this section  we briefly summaries the major equations staring from the EFT of 
 inflation \cite{Bartolo:2015qvr}, a model independent framework developed to analyze primordial fluctuations,
 that will help us calculate the 3-point function for PGW with NBD states. 
 As pointed out earlier, the prospects of EFT of inflation in exploring tensor non-Gaussuainities have been 
 investigated to some extent by the present authors in couple of articles 
 \cite{Naskar:2018rmu,Naskar:2019shl}.
 In this article, our analysis is much more general that takes under consideration both BD and NBD vacua as well as 
 is consistent with Swampland and TCC criteria.

 A consistent EFT of inflation can be given starting from the fact that time diffeomorphism is broken spontaneously 
 after inflation and as a result Goldstone Boson is produced. In unitary gauge the Lagrangian can be written as,
\begin{multline}\label{eq3.1}
	\mathcal{S}=\int d^4x \sqrt{-g}\left[\frac{1}{2}M_{pl}^2R-\Lambda(t)-c(t)g^{00}+
	 \frac{1}{2}M_2(t)^4(g^{00}+1)^2-\frac{\bar{M}_1(t)^3}{2}(g^{00}+1)\delta K_{\mu}^{\mu}
           -\frac{\bar{M}_2(t)^2}{2}\delta K_{\mu}^{\mu2} \right. \\
           \left. 
           -\frac{\bar{M}_3(t)^2}{2}\delta K_{\mu}^{\nu} K_{\nu}^{\mu}\right. 
            \left. +  \frac{M_3(t)^4}{3!}(g^{00}+1)^3-\frac{\bar{M}_4(t)^3}{3!}(g^{00}+1)^2\delta K_{\mu}^{\mu}
            -\frac{\bar{M}_5(t)^2}{3!}(g^{00}+1)\delta K_{\mu}^{\mu 2}\right.\\
            \left.-\frac{\bar{M}_6(t)^2}{3!}(g^{00}+1)
            \delta K_{\mu}^{\nu } \delta K_{\nu}^{\mu}
            -\frac{\bar{M}_7(t)}{3!}\delta K_{\mu}^{\mu 3}
            -\frac{\bar{M}_8(t)}{3!}\delta K_{\mu}^{\mu} \delta K_{\nu}^{\rho} \delta K_{\rho}^{\nu}-
            \frac{\bar{M}_9(t)}{3!}\delta K_{\mu}^{\nu} \delta K_{\nu}^{\rho} \delta K_{\rho}^{\mu}+....\right] 
\end{multline}

Here $\delta K$ is the fluctuation in extrinsic curvature. $M_{i}$ and $\bar{M}_i$ denotes parameters that measures the strength of higher order fluctuations. Different combination of $M_{i}$ and 
$\bar{M}_i$ produce results for different kind of theory. Note  that the scalar 
fluctuation is now represented by the Goldstone Boson $\pi$ produced by broken time diffeomorphism and it 
is related to curvature perturbation $\zeta$ as $\zeta=-H \pi$. 

Another aspect to note here is that the sound speed of scalar and tensor fluctuations get modified due to
the presence of higher order fluctuation term. From \eqref{eq3.1} the second order Lagrangian for 
scalar fluctuation can be written as,
\begin{equation}
	\mathcal{S}^{s}_{2}
   	=\int dt~d^3x~ a^3\left(\frac{\frac{\bar{M_{1}}^3}{2}H-M_{p}^2\dot{H}}{c^2_s}\right)\left[\dot{\pi}^2-c^2_s\frac{(\partial_i \pi)^2}{a^2}\right]
\end{equation}
where scalar sound speed is given by
 \begin{equation}
c_s^2=\frac{\frac{\bar{M_{1}}^3}{2}H-M_{p}^2\dot{H}}{2M_2^4-M_{p}^2\dot{H}}
\end{equation}

Further, from \eqref{eq3.1} the second order tensor fluctuation can be written as,
\begin{equation}
S^{T}_2=\int d^4x \sqrt{-g}\left[\frac{M_{pl}^2}{8}\left(\dot{\gamma}_{ij}^2-
  \frac{(\partial_k \gamma_{ij})^2}{a^2}\right)-\frac{\bar{M}_{3}^{2}}{8}\dot{\gamma}_{ij}^2 \right]
\end{equation}
and the corresponding sound speed of tensor perturbation  is given by
\begin{equation}\label{eq3.5}
c_{t}^2= \frac{M_{pl}^2}{M_{pl}^2-\bar{M}_3^2}
\end{equation}

 As we have seen in the last section, these modified sound speeds of 
perturbations can significantly influence the TCC and Swampland Criteria.
In the rest of the article, we will make use of the above actions as well as the non-trivial sound speeds for scalar and tensor
perturbations to analyse the 3-point statistics for tensor modes using NBD initial state.


\section{The tensor-tensor-tensor ($\gamma \gamma \gamma$) correlator}\label{sec3.2}

Let us begin with  calculating the auto-correlation of tensor modes that gives rise to the
auto-bispectrum, $i,e$ the tensor-tensor-tensor correlator.
To calculate the bispectrum we have to look at the third order Lagrangian of 
tensor fluctuation and the most general Lagrangian can be written as \cite{Gao:2011vs,Naskar:2018rmu},
\begin{equation}\label{eq3.6}
S_{3}^{T}=\int d^4x \sqrt{-g}\left(-\frac{M_{pl}^2}{8}\left(2 \gamma_{ik} \gamma_{jl}-\gamma_{ij}\gamma_{kl} \right)
\frac{\partial_k \partial_l \gamma_{ij}}{a^2}-\frac{\bar{M}_9}{3!} \dot{\gamma}_{ij}  \dot{\gamma}_{jk}  
\dot{\gamma}_{ki}\right)
\end{equation}
In the above action, the term proportional to $M_{pl}$ is the sole contribution  of the so-called Einsteinian or $R$ part and the term proportional to $\bar{M_9}$
is the contribution of higher order gravitational fluctuations. As it will be revealed below, together they play a crucial role in determining the
strength of tensor bispectra.

In order to calculate the 3-point correlation function of tensor fluctuations, we make use of the IN-IN formalism. With the NBD states the 
$\langle\gamma \gamma \gamma\rangle$ 3-point function for any general interaction Hamiltonian
 $H_I(t)$
can be written as,
\begin{equation}
\langle\gamma^{s_1}_{k_1} \gamma^{s_2}_{k_2} \gamma^{s_3}_{k_3}\rangle=-i\int_{\eta_0}^{0} dt'
\langle 0\vert\left[\gamma^{s_1}_{k_1} \gamma^{s_2}_{k_2} \gamma^{s_3}_{k_3}, H_I(t')\right]\vert 0 \rangle
\end{equation}
where $s_i =\{+, \times \}$ are the polarization indices.
An essential point to note here is that   the lower limit of the integration is different from the BD case which can give rise to
a non-trivial term proportional to $e^{i K \eta_0}$, where $K$ can be any possible combination of 
$(k_1\pm k_2\pm k_3)$. However,  since we are working with an initial condition where modes
are within the horizon, $k \eta_0>>1$. This causes a rapid oscillation of the exponential factor and 
 any exponential factor, upon averaging,  will have practically zero contribution. 

 So, with NBD states the contribution of the $R$ part to the $\langle\gamma \gamma \gamma\rangle$ correlator is given by
\begin{multline}\label{eq3.8}
\langle\gamma^{s_1}_{k_1} \gamma^{s_2}_{k_2} \gamma^{s_3}_{k_3}\rangle|_R= \frac{H^4}{32 M_{pl}^4 c_{t}^3}\frac{1}{k_1^3 k_2^3 k_3^3}(s_1 k_1+s_2 k_2+s_3 k_3)^2 F(s_1 k_1,s_2 k_2,s_3 k_3)
(\alpha^t_1+\beta^t_1)(\alpha^t_2+\beta^t_2)(\alpha^t_3+\beta^t_3) \\
\left[\left(-c_t (k_1+k_2+k_3)+c_{t}\frac{k_1^2(k_2+k_3)+k_2 k_3(k2+k3)+k1(k_2^2+4k_2 k_3+k_3^2)}{(k_1+k_2+k_3)^2}\right)\left(\alpha_1^{t*}\alpha_2^{t*}\alpha_3^{t*}-\beta_1^{t*}\beta_2^{t*}\alpha_3^{t*}\right)\right.\\ \left.+ \left(-c_t (-k_1+k_2+k_3)+c_t\frac{k_1^2(k_2+k_3)+k_2 k_3(k_2+k_3)-k_1(k_2^2+4k_2 k_3+k_3^2)}{(-k_1+k_2+k_3)^2}\right)
\left(\alpha_2^{t*}\alpha_3^{t*}\beta_1^{t*}-\beta_1^{t*}\beta_2^{t*}\alpha_3^{t*}\right)\right.\\ \left.
-\left(-c_t (k_1-k_2+k_3)+c_t\frac{k_1^2(k_2-k_3)+k_2 k_3(-k_2+k_3)-k_1(k_2^2-4k_2 k_3+k_3^2)}{(k_1-k_2+k_3)^2}\right)
\left(\alpha_1^{t*}\alpha_3^{t*}\beta_2^{t*}-\beta_1^{t*}\beta_3^{t*}\alpha_2^{t*}\right)\right.\\ \left.
+\left(-c_t (k_1+k_2-k_3)+c_t\frac{k_1^2(k_2-k_3)+k_2 k_3(-k_2+k_3)-k_1(k_2^2-4k_2 k_3+k_3^2)}{(k_1+k_2-k_3)^2}\right)\left(\alpha_1^{t*}\alpha_2^{t*}\beta_3^{t*}-\beta_1^{t*}\beta_2^{t*}\alpha_3^{t*}\right)  \right]\\
+c.c.
\end{multline}
with the following notations: $\alpha_i= \alpha_{k_i}$,
$\beta_i=\beta_{k_i}$ and so on, and the momentum conserving delta function is dropped here.
Also, a $*$ stands for the complex conjugate of $\alpha_i$ and $\beta_i$, as the case may be.
The term $F(x,y,z)$ appears due to the contraction of polarization tensors. 
Written explicitly, it reads,
\begin{equation}
F(x,y,z)=\frac{(x+y+z)^3(x+y-z)(x-y+z)(-x+y+z)}{x^2y^2z^2}
\end{equation}
with $x, y, z$ are, in this particular scenario, given by the combination of $s_i$ and $k_i$ as given in \eqref{eq3.8}.

To calculate the amplitude of bispectrum we will use the parameterisation \eqref{2.6} and \eqref{2.7}. 
$N_k^{(s/t)}$ can be modeled as $N_k^{(s/t)}\sim N_0^{(s/t)} e^{-k^2/(M_{(s/t)\eta_0})^2}$ and also the phase 
$\theta_k^{(s/t)}$ can have explicit momentum dependence. However, as it was found, 
the amplitude of bispectrum will be much larger for a constant phase factor rather than a momentum dependent
phase factor  \cite{Ganc:2011dy}. Further,  if $M_{(s)}$ lies within the observable mode, $i.e$ if the exponential 
factor is non-negligible, then the amplitude of bispectrum also gets suppressed. So, we consider $N_k^{(s/t)}\sim N_0^{(s/t)}$.

Following \cite{Shiraishi:2019yux,Akrami:2019izv}. let us define the amplitude of tensor non-Gaussianity, i.e., the tensor nonlinearity parameter $f_{NL}$   as,
\begin{equation}\label{eq3.10}
f_{NL}^{s_1 s_2 s_3}=\frac{5}{6} \frac{\langle\gamma^{s_1}_{k_1} \gamma^{s_2}_{k_2} \gamma^{s_3}_{k_3}\rangle}{P_{\zeta}(k_1)
P_{\zeta}(k_2)+P_{\zeta}(k_2)P_{\zeta}(k_3)+P_{\zeta}(k_3)P_{\zeta}(k_1)}
\end{equation}
In the limit $N_0^{(s/t)}>>1$,
one finds that $\Gamma_{(s/t)}\sim N_0^{(s/t)}$. It can also be shown that in the limit $N_0^{(s/t)}>>1$, eq
\eqref{eq3.8} can be estimated as, 
\begin{equation}
\langle\gamma^{s_1}_{k_1} \gamma^{s_2}_{k_2} \gamma^{s_3}_{k_3}\rangle|_R\propto \frac{H^4}{32 M_{pl}^4 c_{t}^2}
 (N_0^{(t)})^2 \times f(k_i)
\end{equation}
where $f(k_i)$ encodes all the momentum dependence. The $(N_0^{(t)})^2$ comes from different combinations of
$\alpha_i^t$ and $\beta_i^t$ in Eq \eqref{eq3.8}.

Consequently, from \eqref{eq3.10} and \eqref{eq2.14} the nonlinearity parameter for $\langle\gamma \gamma \gamma\rangle$ correlation can be estimated as,
\begin{equation}
f_{NL}\vert_{\gamma \gamma \gamma (R)}\propto \left(\frac{\epsilon c_s}{c_t}\frac{\Gamma_{(t)}}{\Gamma_{(s)}}\right)^2\sim r^2
\end{equation}

Because of the above definition of $f_{NL}$, it will always
be proportional to $r^2$,
hence, even if we manage to produce a large value for $N_0^{(t)}$ we can not generate a 
large bispectrum from the $R$ part of the Einstein term  alone. 


However, we have in our hand contribution from another parameter,
namely, the $\bar{M}_9$ operator, that arises from the higher order gravitational fluctuations in EFT.
Let us now  discuss the contribution of $\bar{M}_9$ operator to the 3-point tensor correlator. With this term playing the central role,
the 3-point function for the EFT term can be written as,
\begin{multline}
\langle\gamma^{s_1}_{k_1} \gamma^{s_2}_{k_2} \gamma^{s_3}_{k_3}\rangle|_{EFT}=\frac{\bar{M_9} H^5}{M_{pl}^6}(\alpha^t_1+\beta^t_1)(\alpha^t_2+\beta^t_2)(\alpha^t_3+\beta^t_3) F(x,y,z)\\ 
\left[\frac{\left(\alpha_1^{t*}\alpha_2^{t*}\alpha_3^{t*}-\beta_1^{t*}\beta_2^{t*}\beta_3^{t*}\right)}{(k_1+k_2+k_3)^3} -\frac{\left(\alpha_2^{t*}\alpha_3^{t*}\beta_1^{t*}-\beta_2^{t*}\beta_3^{t*}\alpha_1^{t*}\right)}{(k_1-k_2-k_3)^3}+\frac{\left(\alpha_1^{t*}\alpha_3^{t*}\beta_2^{t*}-\beta_1^{t*}\beta_3^{t*}\alpha_2^{t*}\right)}{(k_1-k_2+k_3)^3}+\frac{\left(\alpha_1^{t*}\alpha_2^{t*}\beta_3^{t*}-\beta_1^{t*}\beta_2^{t*}\alpha_3^{t*}\right) }{k_1+k_2-k_3}\right]\\
+c.c.
\end{multline}

The TCC constraint $\epsilon<10^{-31}$ gives the maximum value for $H$ as $H\sim 3^{3/2}\times 10^{-20}M_{pl}$; Consequently, with the current observational 
constraints mentioned earlier, the estimates for different limits of  $f_{NL}$ for the operator 
$\bar{M_9}$ are as under: \\
For equilateral limit $f_{NL}^{eq}\sim 10^{-23}$ whereas
for squeezed limit: $f_{NL}^{sq}\sim 10^{-21}$. These estimations are done under the assumption
that $\bar{M_9}\sim M_{pl}$ and $c_t=1$ to estimate the maximum contribution of this operator.
The details of $f_{NL}$ parameter is given in \ref{appendix}.
This indicates that the contribution of this operator is highly suppressed for TCC as compared to
the result reported in \cite{Naskar:2018rmu}.
From this above discussion we can see that even with NBD
states the signal strength for $\langle\gamma \gamma \gamma\rangle$ correlator arising from the combination
of Einstein $R$ and the EFT term is too feeble to be detected, at least in next generation CMB missions.




\section{The tensor-tensor-scalar ($\gamma \gamma \zeta$) correlator}

As the auto bispectrum, the mixed correlation between scalar and 
tensor modes can also help us to estimate the parameters of PGW. In \cite{Noumi:2014zqa} it is shown that
$\bar{M}_3 \delta K_{\nu}^{\mu}\delta K_{\mu}^{\nu}$ operator can contribute to $\langle  \gamma \gamma \zeta  \rangle$ correlation in decoupling limit and it is very sensitive to the sound speed of tensor fluctuation. 
This is because  \eqref{eq3.5} reveals that the tensor sound speed is modified due to the presence of 
$\bar{M}_3$. So 
it is evident that a 
non zero $\bar{M}_3$ $i,e$ $c_t \neq 1$ will lead to non zero amplitude of $\langle  \gamma \gamma \zeta \rangle$
correlator.

In decoupling limit, the part of the action that is contributing to  the above mixed correlator arises from $\bar{M}_3 \delta K_{\nu}^{\mu}\delta K_{\mu}^{\nu}$ and
can be written as,
\begin{equation}\label{eq3.12}
\mathcal{S}_{\gamma \gamma \zeta}=\frac{\bar{M}_3}{4} \int d^4 x \sqrt{-g} \dot{\gamma}_{ij}
\frac{\partial_k \gamma_{ij}\partial_k \pi}{a^2}
\end{equation}

From the above  action \eqref{eq3.12}, we can calculate the $\langle  \gamma \gamma \zeta  \rangle$ correlator that reads
\begin{multline}\label{stt}
\langle \gamma_{k_2}^{s_2}\gamma_{k_3}^{s_3} \zeta\rangle=-\frac{\bar{M_3}H^4}{16 M_{pl}^6 c_s \epsilon}(\alpha^s_1+\beta^s_1)(\alpha^t_2+\beta^t_2)(\alpha^t_3+\beta^t_3)\epsilon_{ij}^{s_2}(k_2)\epsilon_{ij}^{s_3}(k_3)\frac{(\vec{k_1}.\vec{k_2})}{k_1^3k_2^3k_3}\\
\left[-\frac{2c_s^2 k_1^2+3 c_s c_t k_1(2 k_2+k_3)+c_t^2(k_2+k_3)(2k_2+k_3)}{(c_s k_1+ c_t(k_2+k_3))^3}\left(\alpha_1^{s*}\alpha_2^{t*}\alpha_3^{t*}-\beta_1^{s*}\beta_2^{t*}\alpha_3^{t*}\right)+ \right. \\
\left. \frac{2c_s^2 k_1^2-3 c_s c_t k_1(2 k_2+k_3)+c_t^2(k_2+k_3)(2k_2+k_3)}{(c_s k_1-c_t(k_2+k_3))^3}
\left(\alpha_2^{t*}\alpha_3^{t*}\beta_1^{s*}-\beta_2^{t*}\beta_3^{t*}\alpha_1^{s*}\right)-\right. \\
\left. \frac{2c_s^2 k_1^2-3 c_s c_t k_1(2 k_2+k_3)+c_t^2(-k_2+k_3)(-2k_2+k_3)}{(c_s k_1+c_t(-k_2+k_3))^3}
\left(\alpha_1^{s*}\alpha_3^{t*}\beta_2^{t*}-\beta_1^{s*}\beta_3^{t*}\alpha_2^{t*}\right)-\right. \\
\left. \frac{2c_s^2 k_1^2-3 c_s c_t k_1(2 k_2-k_3)+c_t^2(k_2-k_3)(2k_2-k_3)}{(c_s k_1+c_t(k_2-k_3))^3}
\left(\alpha_1^{s*}\alpha_2^{t*}\beta_3^{t*}-\beta_1^{s*}\beta_2^{t*}\alpha_3^{t*}\right)
 \right]
\\
+k_2 \leftrightarrow k_3+c.c.
\end{multline}
Here, as before, the $s_i$ are the polarization indices and $\alpha_i$ and $\beta_i$ follow the definition as given earlier. The $\epsilon_{ij}^s(k)$'s are the polarization tensors (not to be confused with the first slow roll parameter
$\epsilon$ used in the analysis). Note that these tensors did not appear in the earlier 
correlators as they were completely contracted.
Eq. \eqref{stt} is calculated considering nontrivial sound speeds of scalar and tensor fluctuations. 

However, current 
observations  do not put any constraint on this correlator. So, in what follows we will mostly be interested to find out how much this signal can 
be enhanced with the theoretical constraints from SC and TCC, and whether or not that can be of interest for upcoming CMB missions.

From Eq. \eqref{stt} it transpires that 
both $\beta_k^{(s)}$ and $\beta_k^{(t)}$
contribute to the mixed correlator. As a consequence, both $N_0^{(s/t)}$ and $\theta^{(s/t)}$ would eventually 
contribute to the signal strength of the corresponding mixed bispectrum.
However,  setting  $\theta^{(s)}=\pi$  leads to $N_0^{(s)}>>1$  \cite{Flauger:2013hra}. As we have already
stated, a large $N_0^{(s)}$ leads to large $N_0^{(t)}$ we choose $\theta^{(s)}=\pi$. Further, we can also
write the term $\epsilon_{ij}^{s_2}(k_2)\epsilon_{ij}^{s_3}(k_3)$ as,
\begin{eqnarray}
\epsilon_{ij}^{s_2}(k_2)\epsilon_{ij}^{s_3}(k_3)&=&\frac{1}{2} (1-\cos \phi) ~~~\text{for $s_2=s_3$}\\
\epsilon_{ij}^{s_2}(k_2)\epsilon_{ij}^{s_3}(k_3)&=&\frac{1}{2} (1+\cos \phi) ~~~\text{for $s_2=-s_3$}
\end{eqnarray}

Also  \eqref{eq3.5} gives $\bar{M}_3=(1-c_t^{-2})M_{pl}$ and using \eqref{eq2.14} one can estimate 
the correlator as,
\begin{equation}
\langle \gamma_{k_2}^{s_2}\gamma_{k_3}^{s_3} \zeta \rangle \propto 
(c_t^{-2}-1) \frac{c_s \epsilon}{\gamma_s^2} P_{\zeta}^2 (N_0^{(t)})^2
\end{equation}

Therefore, using the same definition of $f_{NL}$ from \eqref{eq3.10} one can write,
\begin{equation}\label{eq3.19}
f_{NL}\vert_{\gamma \gamma \zeta} \propto (c_t^{-2}-1) \frac{c_s \epsilon}{\Gamma_s^2} (N_0{(t)})^2
\sim r c_t (c_t^{-2}-1) \frac{\Gamma_t}{\Gamma_s} 
\end{equation}

Here we make use of the fact that a large $N_0^{(s)}$ can lead to large $N_0^{(t)}$ and in some region of
parameter space $N_0^{(t)}>N_0^{(s)}$, this is why there is no $N_0^{(s)}$ in the numerator of 
the first relation of \eqref{eq3.19}. In this region of parameter space where $N_0^{(t)}>N_0^{(s)}$
and as a consequence $\Gamma_t > \Gamma_s$, \eqref{eq3.19} suggests
that for NBD state the mixed correlator gets an enhancement in the signal.

 The signal strength of the above nonlinearity parameter for the 
$\langle\gamma \gamma \zeta\rangle$ mixed correlator is enhanced due to the presence of the parameters of EFT and NBD states. However, before that, let us explore another
 nontrivial aspect of the correlator under consideration. We can see from \eqref{stt} that the shape of the signal is changed with
respect to the BD case and 
now it gets peaked at either of the following configurations: $c_t(k_2+k_3)=c_s k_1$, $c_s k_1+c_t k_3=c_t k_2$,
$c_s k_1+c_t k_2=c_t k_3$. These configurations are not strictly the folded limit where, $k_i+k_j=k_l$ but
 is nontrivially modified by the sound speed of perturbations. However, with $c_{s/t} =1$, they get back to the folded limit.

It is however important to note that in \cite{Noumi:2014zqa} the authors used a different route to show that the mixed correlator 
for $c_t\neq 1$ can be larger than mixed correlator generated due to shift and lapse function for 
$c_t=1$. For BD case one can show that
\begin{equation}
\frac{\langle \gamma_{k_2}^{s_2}\gamma_{k_3}^{s_3} \zeta \rangle\vert_{c_t\neq 1}}
{\langle \gamma_{k_2}^{s_2}\gamma_{k_3}^{s_3} \zeta \rangle\vert_{c_t=1}}\sim \frac{1}{\epsilon}
\end{equation}

In the light of SC and TCC we can see that, the mixed correlator for $c_t \neq 1$ is now significantly larger
than the $c_t=1$ case even with BD state. So, in principle, the chance of detection of this correlator should
be much higher if $c_t\neq 1$. As a result, the signal strength is significantly enhanced for a nontrivial tensor sound speed
that arises from the higher curvature terms in EFT of inflation. However, as mentioned earlier,
absence of any observational constraint on this correlator from present observation forbids us to do a numerical estimate of 
the corresponding signal strength.
Even in absence of that, one can claim from the above analysis that   these enhanced signals might  of interest for future CMB missions and the estimates might be more relevant once the constraints come in. 


\section{The tensor-scalar-scalar  ($\gamma \zeta \zeta$) correlator}

Till now we have analyzed the correlators with NBD states using EFT of inflation where only consideration was
that the time diffeomorphism gets broken after the end of inflation. In order to calculate the tensor-scalar-scalar mixed correlator,
we consider a 
scenario 
where spatial diffeomorphism is also broken along with the usual time diffeomorphism breaking. This formalism is more rich in structure
and has enormous potential to give rise to novel features, e.g.,
the blue tilt of tensor power spectrum. Corresponding  EFT for the scenario considering the breaking of space-time diffeomorphism
can be written as follows
 \cite{Bartolo:2015qvr}. The expectation value of the symmetry breaking fields look 
\begin{equation}
\bar{\phi}^0(t)=t ~~ ; ~~
\bar{\phi}^{x^i}=\alpha x^i 
\end{equation}  

The fields $\bar{\phi}^0(t)$ and $\bar{\phi}^{x^i}$ are identified as clock and ruler respectively, during inflation 
\cite{Bartolo:2015qvr} and the parameter $\alpha$ measures the breaking of spatial diffeomorphism.
In this framework, the general Lagrangian can be written as,
\begin{equation}\label{eq3.23}
\mathcal{S}=\int d^4 x \sqrt{-g} \left[\frac{M_{pl}^2}{2} R+ F(X,Y,Z)\right]
\end{equation}
where the parameters $X,Y,Z$ defining $F(X,Y,Z)$, which can be any function of the above quantities respecting the symmetry, are given by
\begin{eqnarray}
X^{00}=g^{\mu \nu} \partial_{\mu} \phi^0 \partial_{\nu} \phi^0\\
Y^{0i}= g^{\mu \nu} \partial_{\mu} \phi^0 \partial_{\nu} \phi^i\\
Z^{ij}= g^{\mu \nu} \partial_{\mu} \phi^i \partial_{\nu} \phi^j
\end{eqnarray}
 
First notice that due to the presence of 
$\alpha$ in the action \eqref{eq3.23}, the expression for  power spectrum of scalar fluctuations  is non trivially modified in this scenario. Written explicitly, 
\begin{equation}
P_{\zeta}=\frac{H^4}{8 \pi c_s (-\bar F_X-\alpha^2 \bar F_Y^2/2a^2)} \Gamma_{s}
\end{equation}

Here, the bar on the quantities represents their values as evaluated in the background and the subscripts
represents derivative with respect to $X$, $Y$ or $Z$. Note that the quantity 
$\bar F_y^2/2a^2$ is slowly varying and can be treated as more or less a constant for our evaluation of observable parameters. 
The parameter $\alpha$  is 
also responsible for a tensor mass $m^2_{\gamma}=\alpha^2 (\bar F_z+\alpha^2 \bar F_{zz})/a^2$; 
but in the  limit $\alpha<<1$ this can be ignored and the power spectrum for tensor modes closely resembles \eqref{eq2.15}
 so far as its numerical values are concerned. 
 
The second  nontrivial effect of 
the action \eqref{eq3.23} appears in the $\langle \gamma \zeta \zeta \rangle$ correlator via the  interaction term 
$\frac{F_Y}{2 a^2}\left(\frac{\gamma_{ij}\partial_i\pi \partial_j \pi}{2 a^2}\right)$.
The  $\langle \gamma \zeta \zeta \rangle$ correlator
describes how much local quadruple will affect the power spectrum of scalar perturbations when a
long wavelength tensor mode correlates with small wavelength scalar mode.
 The importance of this term in the action
  is that the behaviour of $\langle \gamma \zeta \zeta \rangle$  generated from this term
does not satisfy the consistency relation of $\langle \gamma \zeta \zeta \rangle$ correlator and can be 
a distinct signature of spatial diffeomorphism breaking during inflation. 
With NBD state the correlator can be written as,
\begin{small}
\begin{multline}
\langle \gamma^{s}_{k_1} \zeta_{k_2} \zeta_{k_3} \rangle= \frac{H^6}{M_{pl}^2 c_{s}^2}\frac{\alpha^2}{8 \pi c_s (-\bar F_X-\alpha^2 \bar F_Y^2/2a^2)} \frac{F_Y}{2 a^2}
\frac{1}{k_1^3 k_2^3 k_3^3}
\epsilon_{ij}^2(k_1)k_2^{i}k_3^j
(\alpha^t_1+\beta^t_1)(\alpha^s_2+\beta^s_2)(\alpha^s_3+\beta^s_3)\\
\left[\left(-(k_1+c_s k_2+c_s k_3)+\frac{c_s k_1^2(k_2+k_3)+c_s^3 k_2 k_3(k2+k3)+c_s^2 k1(k_2^2+4k_2 k_3+k_3^2)}{(k_1+c_s k_2+c_s k_3)^2}\right)\left(\alpha_1^{t*}\alpha_2^{s*}\alpha_3^{s*}-\beta_1^{t*}\beta_2^{s*}\alpha_3^{s*}\right)\right.\\ \left.+ \left((-k_1+c_s k_2+c_s k_3)+\frac{c_sk_1^2(k_2+k_3)+c_s^3 k_2 k_3(k_2+k_3)-c_s^2 k_1(k_2^2+4k_2 k_3+k_3^2)}{(-k_1+c_s k_2+c_s k_3)^2}\right)
\left(\alpha_2^{s*}\alpha_3^{s*}\beta_1^{t*}-\beta_1^{t*}\beta_2^{s*}\alpha_3^{s*}\right)\right.\\ \left.
-\left(-(k_1-c_s k_2+c_s k_3)+\frac{c_s k_1^2(k_2-k_3)+c_s^3 k_2 k_3(-k_2+k_3)-c_s^2 k_1(k_2^2-4k_2 k_3+k_3^2)}{(k_1-c_s k_2+c_s k_3)^2}\right)
\left(\alpha_1^{t*}\alpha_3^{s*}\beta_2^{s*}-\beta_1^{t*}\beta_3^{s*}\alpha_2^{s*}\right)\right.\\ \left.
+\left(-(k_1+c_s k_2-c_s k_3)+\frac{c_s k_1^2(k_2-k_3)+c_s^3 k_2 k_3(-k_2+k_3)-c_s^2 k_1(k_2^2-4k_2 k_3+k_3^2)}{(k_1+c_s k_2-c_s k_3)^2}\right)\left(\alpha_1^{t*}\alpha_2^{s*}\beta_3^{s*}-\beta_1^{t*}\beta_2^{s*}\alpha_3^{s*}\right)  \right]\\
+k_2\rightarrow k_3 +c.c.
\end{multline}	
\end{small}

In squeezed limit $k_1<<k_2,k_3$ if one chooses $\theta^{(s)}=\pi$ as required for large $\beta^s$ and 
$\theta^{(t)}=0$ one finds that,
\begin{equation}\label{eq3.29}
\langle \gamma^{s}_{k_1} \zeta_{k_2} \zeta_{k_3} \rangle \sim \frac{1}{k_1^3 k_2^3}\frac{k_2}{k_1} N^{(s)}_0+
\frac{1}{k_1^3 k_2^3} N^{(t)}_0
\end{equation}

The interesting point to note here is that the second term in \eqref{eq3.29} can be produced with a BD state, 
but the first term is a unique contribution of NBD state which gets an enhancement in squeezed limit of 
$\mathcal{O}(\frac{k_2}{k_1})$. Also it is important to note that for $\theta^{(s)}=\pi$ choice the 
$(N_0^{(s)})^2$ and $N_0^{(s)}N_0^{(t)}$ terms  in the correlator vanish. 
For the first term of \eqref{eq3.29} the expression for the mixed correlator in squeezed limit can be rewritten using the expressions for scalar and tensor power spectra in NBD state, as follows
\begin{equation}\label{eq6.9}
\langle \gamma^{s}_{k_1} \zeta_{k_2} \zeta_{k_3} \rangle \sim P_{\zeta} P_{\gamma} 
\frac{1}{\Gamma_t} \frac{k_2}{k_1}
\end{equation}

It appears from 
the above expression that in order to obtain an enhancement of signal for $f_{NL}$ for this correlator, 
the following condition has to be  satisfed: $\frac{1}{\Gamma_t} \frac{k_2}{k_1}>1$. However, as we have mentioned before,  
TCC requires a large $\Gamma_t\sim 10^{29}$ in order to have a detectable signal of PGW and thus suppressing 
this effect. On the other hand if we are not strict about the theoretical effect on $\theta^{(s)}$ and set 
it any value other than $\pi$ we can have, 
\begin{equation}\label{eq6.10}
\langle \gamma^{s}_{k_1} \zeta_{k_2} \zeta_{k_3} \rangle \sim P_{\zeta} P_{\gamma} 
\frac{\Gamma_s}{\Gamma_t} \frac{k_2}{k_1}
\end{equation}


From \eqref{eq6.10} it is evident that in order to have a large signal for 
this mixed bispectrum, the parameters need to satisfy the following condition: $\frac{\Gamma_s}{\Gamma_t} \frac{k_2}{k_1}>1$ 
which is possible in some region of parameter space  even for large $\Gamma_s$ and $\Gamma_t$. Thus, a considerably strong signal for this correlator is
achievable  with NBD initial state for both scalar and tensor modes along with non-trivial sound speeds, satisfying both SC and TCC. The non-linearity parameter can be defined for this correlator as,

\begin{equation}
f_{NL}\vert_{\gamma \zeta \zeta}= \frac{\langle\gamma^{s} \zeta \zeta\rangle}{P_{\zeta} P_{\gamma}}
\end{equation}

It is easy to see from \eqref{eq6.9} and \eqref{eq6.10} that 
$f_{NL}\vert_{\gamma \zeta \zeta}\propto \frac{1}{\Gamma_t}\frac{k_s}{k_l}$ for $\theta^{(s)}= \pi$ 
and $f_{NL}\vert_{\gamma \zeta \zeta}\propto \frac{\Gamma_s}{\Gamma_t} \frac{k_2}{k_1}$ for 
$\theta^{(s)}\neq \pi$.
It worths mentioning that this enhancement in the mixed correlator  is reflected on the quadrupole moment and 
might be of interest for future CMB missions. This makes the present investigation for this correlator relevant  from the point of view of observations.

\section{Summary and outlook}\label{sec4}

Recently proposed SC and TCC  put stringent theoretical constraints on the amplitude of 2-point correlation function of PGW - the so-called tensor-to-scalar ratio  $r$. 
TCC bound reveals that with the usual BD vacuum, a detectable PGW signal can not be produced from vacuum fluctuation during 
inflation. However, subsequently, it was found that if the initial states are NBD both for scalar and tensor modes,
one can relax the
bound on $r$ sufficiently bringing it back to the observational limit of next generation CMB missions with the help of the NBD parameters
 $\Gamma_s$ and $\Gamma_t$ 
(functions of scalar and tensor Bogolyubov coefficients, respectively).
 In this article we  explored  the 3-point statistics of tensor modes respecting both the constraints coming from SC and TCC, thereby  
considering NBD states for both scalar and tensor modes.
Using  a
model-independent framework of EFT of inflation that helps us keep the analysis and results more or less generic,
 we calculated all possible correlates (auto and mixed) related to tensor non-Gaussianities, followed by
 proposing the possible templates for the non-linearity parameters $f_{NL}$ for different relevant shapes. 
 We also tried to explore if at all any of the bispectra could be observationally relevant for future CMB missions by 
 simply calculating the possible upper bounds for each by tuning the  parameters under consideration.
 
 Our analysis reveals that the auto correlator $\langle \gamma \gamma \gamma \rangle$
 generated from Einstein term $R$ does not get significantly enhanced even with a choice of large Bogolyubov coefficient
 $\beta^{(s/t)}$, since the
definition of the amplitude of bispectrum always keeps it proportional to $r^2$. The  amplitude of the 
auto correlator due to the higher derivative EFT operator too is highly suppressed when one considers TCC. 
Thus,  the prospects of detecting the tensor auto corrector are almost nil, at least in next generation CMB missions.
We have also analyzed the mixed $\langle  \gamma \gamma   \zeta \rangle$ 
 correlator  which is a good probe of sound speed of PGW. We have found that 
the shape of the correlator is modified due to the NBD state and gets peaked at folded limit modified by 
the sound speed of perturbation. Also the amplitude of this correlator can be estimated as 
$r \frac{\Gamma_t}{\Gamma_s}$ which can be large in the large $\Gamma_s$ limit. 
Thus, this correlator can be relevant for future CMB missions. 
The behaviour of the other mixed correlator, namely, the
$\langle \gamma \zeta \zeta \rangle$ generated due to space-time diffeomorphism breaking is also explored. We have
used the approach of EFT of space-time diffeomorphism breaking for this scenario. This type of correlator 
can be a probe of spatial diffeomorphism breaking during inflation. We have found that using NBD state 
there will be an extra contribution on top of the BD contribution. In squeezed limit this contribution is 
proportional to $\frac{k_s}{k_l}$, where $k_s$  and $k_l$ are respectively the short and the long wavelength that can be probed by
any CMB mission under consideration. The amplitude
of this signal is highly dependent on the phase factor given by the $\theta^{(s)}$ parameter. For any choice of $\theta^{(s)}\neq \pi$,
the amplitude of  this correlator can be significantly enhanced, thereby making it relevant to future CMB missions. 

In a nutshell, if one begins with the theoretical constraints in the light of SC and TCC 
 for slow roll, single field inflation,  the nonlinearity parameter $f_{NL}$ for tensor auto correlation would still be undetectable whereas the corresponding 
 parameters for mixed correlations might be a point of interest for future CMB missions. Our results are more or less generic as we investigated for he parameters from a model-independent  framework of EFT.
 
 We must, however, admit that our analysis in this article is mostly theoretical and in order to find out the prospects of tensor non-Gaussianities in CMB, we made use of the simplistic numerical calculations based on current bounds given by Planck 2018 as well as the future goals of next generation missions like 
 CMB-S4, LiteBIRD and COrE. A detailed forecast with particular mission is required to analyze the finer details of the results as well as
 to estimate the parameters with error in an observation-first approach. Such an analysis is beyond the scope of the present paper. We hope to be 
 back with that in near future.

\section*{Acknowledgments}

AN thanks ISI Kolkata for financial support through Senior Research Fellowship.
\begin{appendix}
\section*{Appendix}
\label{appendix}

Here we introduce the expression for $f_{NL}$ for the auto correlator generated from $R$ term and 
higher order EFT term proportional to $\bar{M}_9$
in $N_0^{(t)}\gg 1$ limit. Here we only give the expression for $+++$ polarisation 
combination.

The expression for $f_{NL}$ in the equilateral limit $R$ term is given as,

\begin{equation}
f_{NL}^{+++,eq}\vert_{R}=\frac{855}{256} r^2 \left(\frac{k}{k_*}\right)^{-2(n_s-1)}
\end{equation}

Where $n_s$ is the spectral tilt.

In the squeezed limit it can be written as,

\begin{equation}
f_{NL}^{+++,sq}\vert_{R}= \frac{5}{18} r^2 \frac{k_s}{k_l} \left(\frac{k_l}{k_*}\right)^{-2(n_s-1)}
\end{equation}

Next we discuss the contribution of the operator proportional to $\bar{M}_9$. The parameter 
$f_{NL}$ for this operator can be written as,

\begin{equation}
f_{NL}^{+++,eq}\vert_{EFT}=\frac{35}{6}\frac{\bar{M}_9 H}{M_{pl}^2} r^2 c_t^2 \left(\frac{k}{k_*}\right)^{-2(n_s-1)}
\end{equation}

And in squeezed limit the contribution becomes,

\begin{equation}
f_{NL}^{+++,sq}\vert_{EFT}=\frac{10}{9}\frac{\bar{M}_9 H}{M_{pl}^2} r^2 c_t^2 \frac{k_s}{k_l}
 \left(\frac{k_l}{k_*}\right)^{-2(n_s-1)}
\end{equation}

\end{appendix}

\end{document}